\documentclass[prl,aps,nofootinbib,notitlepage,twocolumn]{revtex4-1}

\usepackage{amsmath}
\usepackage{amssymb}
\usepackage{graphicx}

\usepackage{txfonts}
\usepackage{dsfont}

\usepackage[colorlinks=true,linkcolor=blue,citecolor=blue,urlcolor=blue]{hyperref}

\newcommand{\<}{\langle}
\renewcommand{\>}{\rangle}
\providecommand{\abs}[1]{\left\lvert#1\right\rvert}

\providecommand{\tr}{{\rm tr}}
\renewcommand{\phi}{\varphi}

\begin{document}
\title{Tailoring Non-Gaussian Continuous-Variable Graph States}

\author{Mattia Walschaers} 
\email{mattia.walschaers@lkb.upmc.fr}
\affiliation{Laboratoire Kastler Brossel, Sorbonne Universit\'e, CNRS, ENS-PSL Research University, Coll\`ege de France; 4 place Jussieu, F-75252 Paris, France}
\author{Supratik Sarkar}
\affiliation{Laboratoire Kastler Brossel, Sorbonne Universit\'e, CNRS, ENS-PSL Research University, Coll\`ege de France; 4 place Jussieu, F-75252 Paris, France}
\author{Valentina Parigi}
\affiliation{Laboratoire Kastler Brossel, Sorbonne Universit\'e, CNRS, ENS-PSL Research University, Coll\`ege de France; 4 place Jussieu, F-75252 Paris, France}
\author{Nicolas Treps}
\email{nicolas.treps@lkb.upmc.fr}
\affiliation{Laboratoire Kastler Brossel, Sorbonne Universit\'e, CNRS, ENS-PSL Research University, Coll\`ege de France; 4 place Jussieu, F-75252 Paris, France}

\date{\today}

\begin{abstract}
Graph states are the backbone of measurement-based continuous-variable quantum computation. However, experimental realisations of these states induce Gaussian measurement statistics for the field quadratures, which poses a barrier to obtain a genuine quantum advantage. In this letter, we propose mode-selective photon addition and subtraction as viable and experimentally feasible pathways to introduce non-Gaussian features in such continuous-variable graph states. In particular, we investigate how the non-Gaussian properties spread among the vertices of the graph, which allows us to show the degree of control that is achievable in this approach. 
\end{abstract}

\maketitle

The quest towards new quantum technologies explores a wide range of potential paths, with the optimal route yet left to be determined. Our present work approaches this problem from the perspective  of {\em continuous variables} (CV) \cite{braunstein_quantum_2005}. CV quantum optics offers several advantages for information processing, such as resilience against decoherence effects and an infinitely large Hilbert space to harbour information. Nevertheless, the most remarkable of these advantages is the capability to deterministically generate large entangled states \cite{PhysRevLett.112.120505,Su:12,PhysRevLett.114.050501,yoshikawa_invited_2016}. This large-scale entanglement can in turn be employed to tailor CV graph states \cite{PhysRevA.83.042335} which form the backbone of measurement-based quantum computation \cite{gu_quantum_2009}. Finally, this approach has become a viable pathway to quantum computation since it was shown that realistic squeezing levels are sufficient to reach fault-tolerance \cite{menicucci_fault-tolerant_2014}, with 10dB as current threshold \cite{PhysRevX.8.021054}.

CV quantum optics relies on the measurement of the field quadratures of  light, i.e.~the real and imaginary parts of the amplitude of the electric field. The CV graph states that can be deterministically generated are Gaussian states, and thus induce Gaussian statistics for quadrature measurements.  However, Gaussian statistics is efficiently simulated classically \cite{bartlett_efficient_2002,mari_positive_2012,rahimi-keshari_sufficient_2016}. With only CV graph states and quadrature measurements, a genuine quantum advantage is unachievable. Thus, methods must be devised to induce much more intricate non-Gaussian statistics in these quantum states in order to develop a feasible platform for universal quantum computation. The theoretically appealing proposal to use a cubic phase gate for inducing non-Gaussian statistics in measurement-based CV quantum computation is experimentally challenging \cite{PhysRevA.84.053802,PhysRevA.91.032321,PhysRevA.93.022301,arzani_polynomial_2017}, which is a key motivation to explore alternative routes.
 
In quantum optics photon subtraction \cite{ourjoumtsev_increasing_2007,parigi_probing_2007,ra_tomography_2017} and addition \cite{zavatta_experimental_2007} provide an experimentally feasible alternative to introduce non-Gaussian features in quantum states of light. Recently, setups for mode-tuneable coherent photon subtraction from multimode light have been tested \cite{PhysRevA.89.063808,averchenko_multimode_2016,ra_tomography_2017}, which led us to develop a general theoretical framework \cite{walschaers_entanglement_2017, walschaers_statistical_2017} to describe the resulting non-Gaussian states. This framework allows us to uncover a wide range of quantum properties of single-photon added and subtracted states through their multimode Wigner functions. Rather than focusing on properties that are specific to the topology of the graph state to (from) which a photon is added (subtracted), we ask the general question how non-Gaussian features spread through a graph state and how well they can be controlled. We will show that the effect of local photon addition or subtraction remains {\em local}, such that non-Gaussian features can be tailored. Here, local is understood with respect to the graph's topology: the effect of photon addition or subtraction in a single vertex spreads up to its {\em next-to-nearest neighbours}. On the other hand, for photon addition or subtraction in a superposition of vertices, we will highlight a trade-off between the strength and the spread of non-Gaussian properties.\\
 
 
Let us start by introducing the system. The CV approach to quantum optics is based on the analysis of the {\em quadratures} of the electromagnetic field. More specifically, it is common to use the complex representation of the electric field operator $\hat{E}(\mathbf{r},t)$ in terms of a basis $\{u_1({\bf r}, t), \dots, u_m({\bf r},t)\}$ of $m$ normalised modes: \footnote{These $u_j({\bf r}, t)$ are normalised solutions to Maxwell's equations. The normalisation is considered with respect to the spacial degrees of freedom, i.e.~$\int {\rm d}^3 r\, \abs{u_j({\bf r},t)}^2  = 1$ for every time $t$.}
\begin{equation}
\hat{E}(\mathbf{r},t) = \sum_{j=1}^m (\hat{x}_j + i \hat{p}_j)u_j({\bf r}, t),
\end{equation}
where $\hat{x}_j$ and $\hat{p}_j$ are the quantum observables for the {\em amplitude and phase quadratures}, respectively. Thus, they obey the canonical commutation relations $[\hat{x}_j, \hat{p}_k] = 2i \delta_{j,k}$, $[\hat{x}_j, \hat{x}_k] = 0$, and $[\hat{p}_j, \hat{p}_k] =0$. 

To connect the quadrature operators to the eventual measurement statistics, one requires a description of the quantum state of the light. In optics, it is common to represent the state by a multimode quasi-probability distribution on the phase space \cite{cahill_density_1969}. In quantum statistical mechanics, one commonly finds a complementary approach, based on the state's correlation functions \cite{verbeure_many-body_2011}. When the measurement statistics for the field quadratures is Gaussian, a characterisation through correlations is particularly interesting, since only those between pairs of quadrature operators (contained in its covariance matrix $V$) are required.

Because CV graph states also belong to the class of Gaussian states, they too can be characterised entirely through their covariance matrix. In the idealised case \cite{gu_quantum_2009}, the modes on which the graph is built are infinitely squeezed, which is unphysical. However, it has been shown that \cite{menicucci_fault-tolerant_2014,su_2018} the quantum computation framework can be adjusted to deal with finite amounts of squeezing. Here, we focus on the more physical situation where squeezing values are assumed to be finite.

To construct an $m$-mode graph state, we start out from a set of independent squeezed modes, with a joint covariance matrix given by 
$V_0={\rm diag}(s_1,\dots, s_m, s_1^{-1}, \dots, s_m^{-1}).$ This collection of non-correlated modes is turned into a graph state by applying entangling operations that are represented by the {\em edges} of the graph. More specifically, every edge of the graph corresponds to a $C_Z$ gate, i.e.~an application of the unitary operator $\exp(i \hat{x}_j\hat{x}_k)$. On the level of covariance matrices, the action of the set of $C_Z$ gates is represented by a symplectic transformation $G$, applied to $V_0$:
\begin{equation}\label{eq:clusterV}
V_0 \mapsto V = G^tV_0G, \text{ with } G= \begin{pmatrix} \mathds{1}& {\cal A}\\ 0  & \mathds{1}\end{pmatrix},
\end{equation}
where $V$ is the $2m$-dimensional covariance matrix of the CV graph state, described by the graph with $m$-dimensional adjacency matrix ${\cal A}$. The covariance matrix $V$ now describes a set of {\em entangled modes}, which we will refer to as the {\em vertices} of the graph.


We can then introduce non-Gaussian features in these states by addition or subtraction of a photon in one of the vertices of the graph (i.e.~one of the initially squeezed modes). To experimentally implement this operation in a fully coherent way, nonlinear optics is required \cite{PhysRevA.89.063808,ra_tomography_2017}. In theory, however, such an operation is accurately modeled via the action of a creation or annihilation operator on the state \cite{averchenko_multimode_2016}. Note, however, that this is not a unitary operation, such that we must renormalise the state after the creation or annihilation operator has acted. This renormalisation factor is given by the success probability of the operation, and reflects the probabilistic nature of the process. The experimental implication of this non-unitarity, is the need for post-selection.

To theoretically describe the resulting non-Gaussian states, we resort to the {\em Wigner function} \cite{HILLERY1984121} as a preferred phase space representation. Wigner functions are of particular interest due to their experimental connection to homodyne tomography \cite{lvovsky_continuous-variable_2009}. From \cite{walschaers_entanglement_2017, walschaers_statistical_2017} we directly obtain the Wigner function for any set of modes in the photon-added or -subtracted graph state. The key ingredients in the theoretical construction are the covariance matrix $V$ of the initial graph state, and the matrix $A_j$ that describes all non-Gaussian effects induced through the addition (``+'') or subtraction (``-'') of a single photon in the $j$th vertex:\footnote{Note that the framework of \cite{walschaers_entanglement_2017, walschaers_statistical_2017} is much more general and allows for photon addition and subtraction in any mode $g \in \mathcal{N}(\mathbb{R}^{2m})$. In this manuscript we focus only on subtraction from the $j$th vertex in the graph state, hence we choose $g$ a vector with components $g_k = \delta_{j,k}$. }
\begin{equation}\label{eq:A}
A_j = 2 \frac{(V \pm \mathds{1}) (P_{j} + P_{j+m})(V \pm \mathds{1})}{{\rm tr} [(V \pm \mathds{1}) (P_{j} + P_{j+m})]}, 
\end{equation}
where $P_j$ denotes the matrix with components $\{P_j\}_{kl} = \delta_{kj}\delta_{lj}$, such that $P_{j} + P_{j+m}$ is the projector on the two-dimensional phase space associated with the $j$th vertex in the graph state.\footnote{$P_j$ projects on the phase space axis associated with the mode's amplitude quadrature, whereas $P_{j+m}$ does the same for the phase quadrature.} The Wigner function for a subset of vertices of the graph, with labels ${\cal V} = \{k_1, \dots k_n\}$, is then given by 
\begin{align}\label{eq:redWigner}
W_{\cal V}(\beta)= \frac{1}{ 2}& \big[(\beta, {V_{\cal V}}^{-1} {A_{j}}_{\cal V} {V_{\cal V}}^{-1} \beta') -  \tr\{{V_{\cal V}}^{-1}{A_{j}}_{\cal V}\} + 2 \big]\\
&\times  \frac{e^{-\frac{1}{2}(\beta, {V_{\cal V}}^{-1} \beta)}}{(2\pi)^n \sqrt{\det V_{\cal V}}},\nonumber
\end{align}
where $\beta = (x_1,\dots, x_n,p_1, \dots, p_n) \in \mathbb{R}^{2n}$ denotes a point in the phase space that is associated with the vertices ${\cal V}$. $V_{\cal V}$ and ${A_{j}}_{\cal V}$ are $2n$-dimensional sub-matrices of $V$ and $A_{j}$, respectively, for the subset of vertices ${\cal V}$.\footnote{As a simple example, note that $$V_{\{k\}} = \begin{pmatrix} V_{k\,k} & V_{k\,k+m} \\ V_{k+m \, k} & V_{k+m\, k+m} \end{pmatrix}. $$} The Wigner function for the $k$th vertex is then obtained by setting ${\cal V} = \{k\}$, whereas the Wigner function for the entire graph is obtained for ${\cal V} = \{1, \dots, m\}$.

The purity of the initial Gaussian graph state implies that the full photon-added or subtracted state remains pure. Furthermore, it is straightforward to verify \cite{walschaers_entanglement_2017} that the Wigner function for any full photon-added and subtracted graph states (i.e.~for all $m$ vertices) is negative. However, for a subset ${\cal V}$ of vertices, the entanglement in the system significantly reduces the purity of reduced state's Wigner function, such that it is no longer guaranteed to have a negative region in phase space. To assess the presence of such a negative region, it suffices \cite{walschaers_statistical_2017} to check the condition $\tr\{{V_{\cal V}}^{-1}{A_{j}}_{\cal V}\} > 2$.\\

As a key result of this letter, we investigate the spread of non-Gaussian features through the graph. In this regard, it is important to understand that the non-Gaussian part of the Wigner function (\ref{eq:redWigner}) is governed by the matrix ${A_{j}}_{\cal V}$. In particular, we recover the Wigner function for a Gaussian graph state whenever ${A_{j}}_{\cal V}=0$. In other words, any subgraph on the set of vertices ${\cal V}$ is unaffected by photon addition or subtraction whenever ${A_{j}}_{\cal V}=0$, and thus we explore the components of $A_j$ as determined by (\ref{eq:A}). 

To do so, we introduce the vector $(0, \dots 0,1,0,\dots,0)^t \equiv e_i \in \mathbb{R}^{2m}$, where the one occurs on the $i$th position. From (\ref{eq:A}) it directly follows that the components of $A_j$ are fully determined by the following identities:
\begin{align}
(e_k, [V \pm \mathds{1}] e_j) &= (G e_k, V_0 G e_j ) \pm (e_k,e_j) \label{thisone}\\
&= (s_k \pm 1) \delta_{j,k}, \nonumber\\
(e_k, [V \pm \mathds{1}] e_{j+m}) &= s_j^{-1} {\cal A}_{jk},\\
(e_{k+m}, [V \pm \mathds{1}] e_{j}) &= s_k^{-1} {\cal A}_{jk},\\
(e_{k+m}, [V \pm \mathds{1}] e_{j+m}) &= (s_{k}^{-1} \pm 1)\delta_{j,k}  + \sum_{l=1}^m s_k^{-1} {\cal A}_{kl}{\cal A}_{jl}.\label{thatone}
\end{align}
These identities are only non-zero when $k = j$, when $k \in N(\{j\})$, or when $k \in N(N(\{j\}))$, where $N({\cal X})$ is the {\em neighbourhood} of the set of vertex labels ${\cal X}$, i.e.~$N({\cal X}) \equiv \big\{k \in \{1, \dots m\} \mid k\notin {\cal X}, \exists l \in {\cal X}: {\cal A}_{kl} = 1 \big\}$. It follows that the Wigner function $W_{\cal V}(\beta)$ for any subgraph that does not contain vertices that are two steps or less removed from the vertex $j$ is Gaussian and completely unaltered by the addition or subtraction of the photon in $j$. Formally, we can express this main result of our work as 
\begin{equation}\begin{split}\label{eq:Main}
{\cal V} \cap \Big(\{j\} \cup N(\{j\}) \cup N(N(\{j\}))\Big) = \emptyset \\
\implies W_{\cal V}(\beta)= \frac{e^{-\frac{1}{2}(\beta, {V_{\cal V}}^{-1} \beta)}}{(2\pi)^n \sqrt{\det V_{\cal V}}},
\end{split}
\end{equation}
This implies that {\em non-Gaussian properties, induced by photon addition and subtraction, spread over a maximal distance of two vertices.}\footnote{Note that the proof relies only on the fact that $V_0$ is a collection of uncorrelated modes, but not on the purity of the state of each of these modes.}\\

\begin{figure}
\centering
\includegraphics[width=0.4\textwidth]{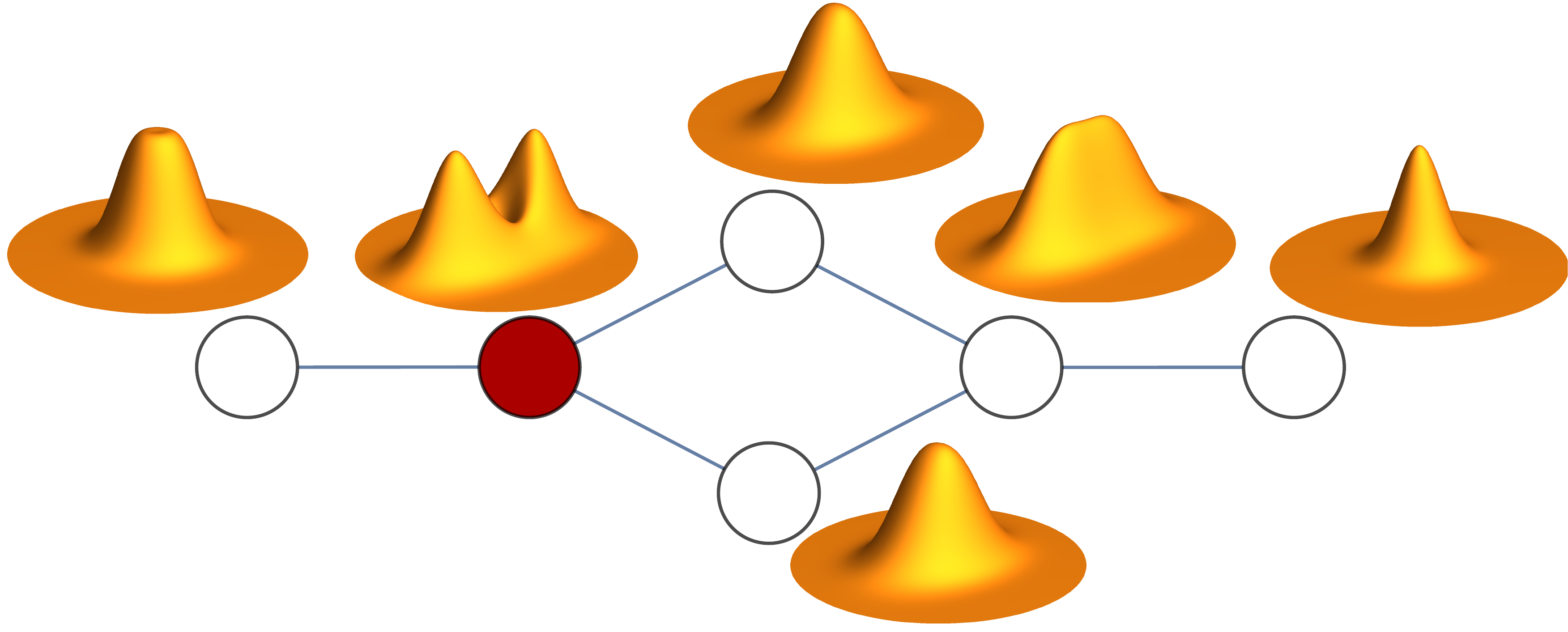}
\caption{Graph state (\ref{eq:clusterV}) of six vertices (modes), with a photon subtracted in the mode associated with the red vertex. The single-mode Wigner function (\ref{eq:redWigner}) with ${\cal V} = \{k\}$ is shown for each vertex $k$. All modes in the initial squeezed vacuum $V_0$ are equally squeezed (i.e.~$s_1= \dots =s_m$) at $10{\rm dB}$.\label{fig:GraphWig}}
\end{figure}

We now show graphically examples of how non-Gaussian properties manifest within the system, representing the Wigner function (\ref{eq:redWigner}) of individual vertices. In Fig.~\ref{fig:GraphWig} we show this for photon subtraction from a particular six-mode graph state where all initial modes in $V_0$ are equally squeezed. For every vertex $k$, we show the associated Wigner function that is obtained by setting ${\cal V} = \{k\}$ in (\ref{eq:redWigner}). Single mode Wigner functions are important tools to gain an intuition in the system, as they can convey qualitative properties of the state. As such, we observe that the non-Gaussian effects are most pronounced in the Wigner function of the vertex where the photon was subtracted. Due to the entanglement between the different vertices, the effects of photon subtraction are propelled through the graph. In particular, we see pronounced non-Gaussian features two vertices away from the point of subtraction. However, the Wigner function of the rightmost vertex is Gaussian, and is, thus, unaffected by the subtraction of the photon, as was proven in (\ref{eq:Main}).

\begin{figure}
\centering
\includegraphics[width=0.49\textwidth]{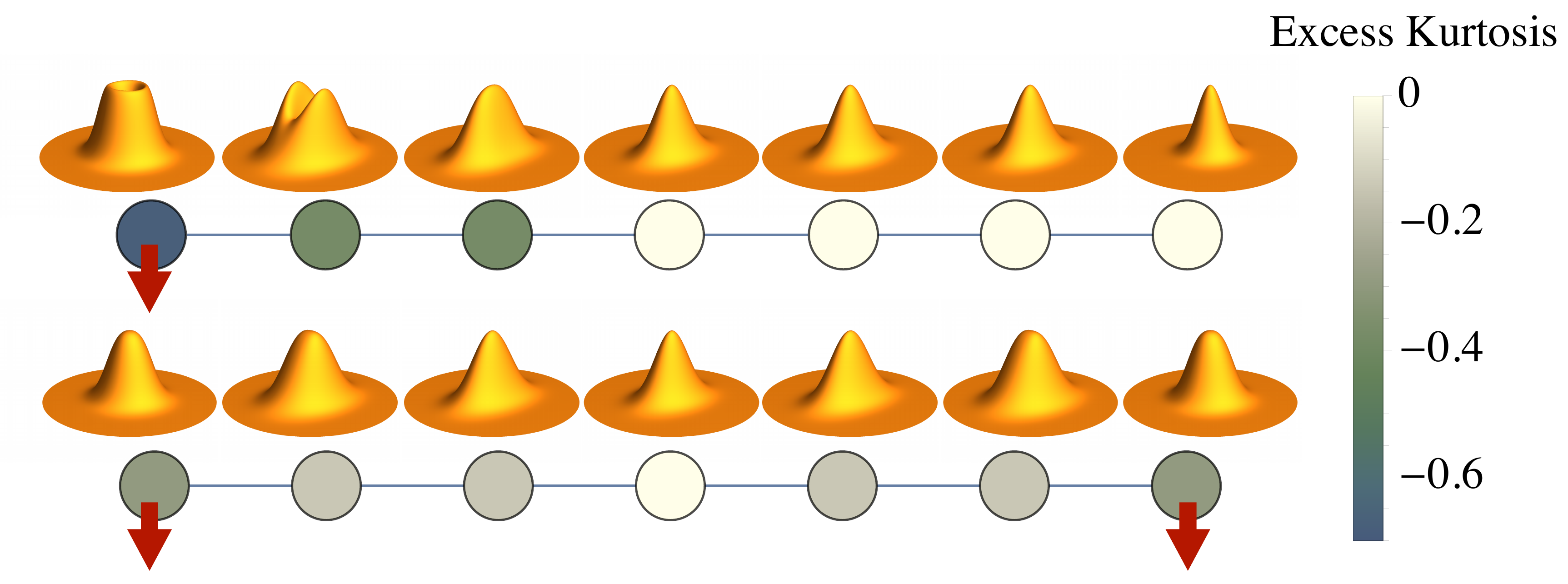}
\caption{Graph state (\ref{eq:clusterV}) of seven vertices (modes), with a photon subtracted in a balanced superposition of modes associated with the vertices that are indicated by a read arrow. The single-mode Wigner function (\ref{eq:redWigner}) with ${\cal V} = \{k\}$ is shown for each vertex $k$, together with the associated excess kurtosis of the phase quadrature (represented in colour code).  All modes in the initial squeezed vacuum $V_0$ are equally squeezed (i.e.~$s_1= \dots =s_m$) at $10{\rm dB}$. \label{fig:3}}
\end{figure}

Note that our result (\ref{eq:Main}) only holds for addition and subtraction in a single vertex. The theoretical framework of \cite{walschaers_entanglement_2017, walschaers_statistical_2017}, and the experimental setup of \cite{ra_tomography_2017} also allow for mode-selective subtraction in a superposition of vertices. Formally, the subtraction in a superposition of vertices $j$ and $k$ can simply be modelled by the action of a superposition of annihilation operators $(a_j + a_k)/\sqrt{2}$ on the state. The result (\ref{eq:Main}) can straightforwardly be extended to incorporate this possibility.\footnote{Equations (\ref{thisone}-\ref{thatone}) can be evaluated for all the vertices $j$ that contribute to the superposition that describes the mode where the photon is added or subtracted. This implies that only vertices that are up to two steps away from one of the vertices in the superposition can be affected by the addition or subtraction of a photon.} We illustrate in Fig.~\ref{fig:3} that such superpositions allow us to spread out non-Gaussian effects over a larger part of the system, in this particular case for a small graph of seven vertices. In the bottom panel, where we subtract a photon from balanced superposition of the first and last vertex, the single-mode Wigner functions up to two steps away from both of these vertices show non-Gaussian properties. However, in the upper panel, where a photon is subtracted purely from the leftmost vertex, we see that the non-Gaussian effects are far more pronounced. To quantify this observation, we resort to the excess kurtosis $\kappa_k=\<\hat{p}_{k}^4\>/\<\hat{p}_{k}^2\>^2-3$ as a hallmark of non-Gaussian measurement statistics for a single quadrature (in this case the phase quadrature). The excess kurtosis is narrowly related to the fourth cumulant, which the first cumulant that clearly shows the non-Gaussian features of the state. For Gaussian statistics, all cumulants beyond the second order are exactly equal to zero. From the correlation functions that were determined in \cite{walschaers_statistical_2017}, we find that for photon-added and subtracted Gaussian states $\kappa_k \leqslant 0$. Any instance where the inequality is strict indicates sub-Gaussian measurement statistics, which implies lighter tails in the probability distribution of measurement outcomes \cite{Westfall:2014aa}.

The results in Fig.~\ref{fig:3} suggest that for the addition or subtraction of a single photon there is a trade-off between the local strength of non-Gaussian features and the spread through the graph. In other words, we can either induce strong local, or a weaker more spread out non-Gaussian effect, depending on how the photon is subtracted.\\

\begin{figure}
\centering
\includegraphics[width=0.35\textwidth]{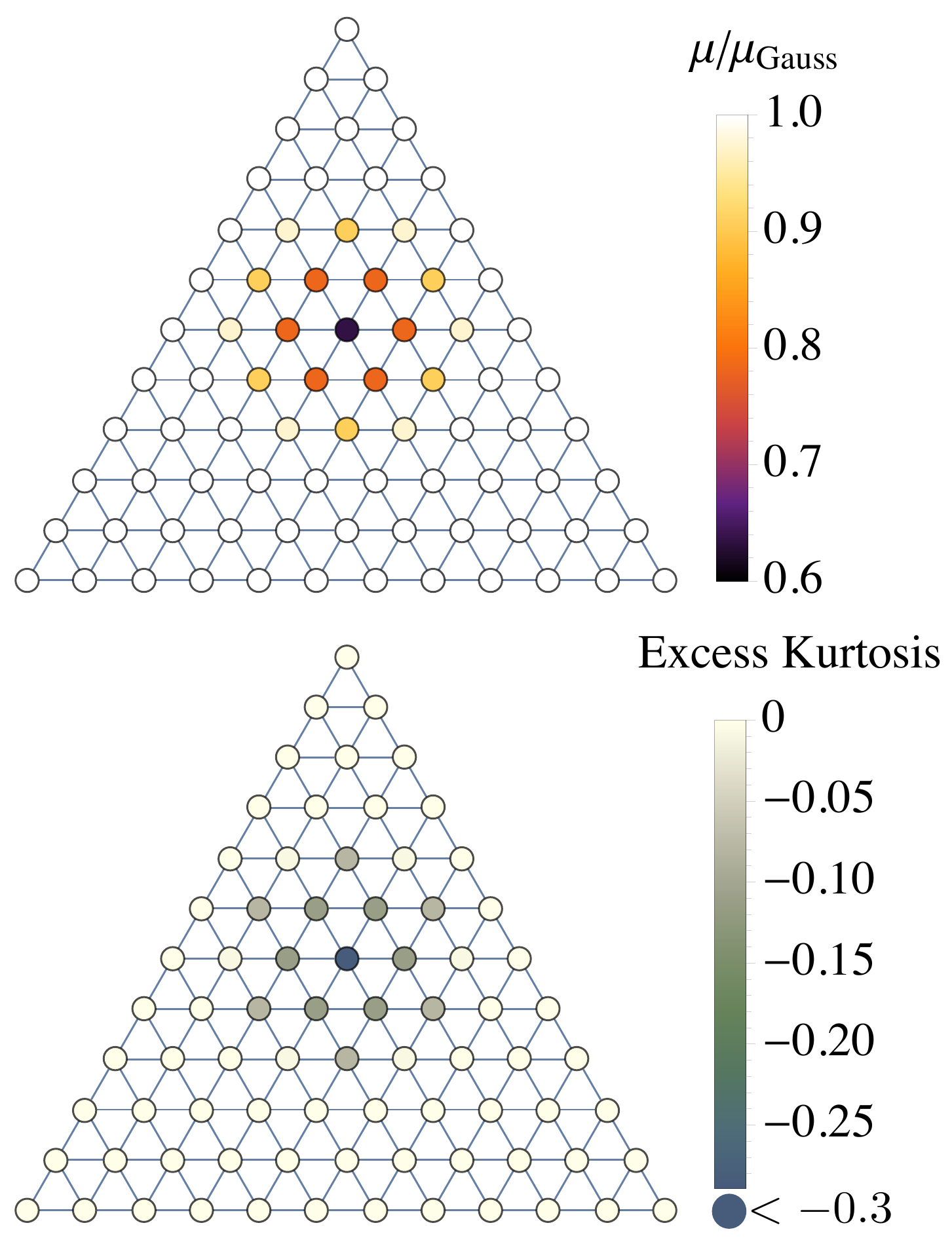}
\caption{Decrease in purity (top) and excess kurtosis (bottom) for every vertex' single-mode Wigner function (\ref{eq:redWigner}), upon subtraction in the centre vertex, are indicated by colour code. Decrease of the purity $\mu$ (\ref{eq:purity}) of the single-mode Wigner function as compared to the same vertex' purity $\mu_{\rm Gauss}$ prior to photon subtraction represents indicates an increase of entanglement between the vertex and the rest of the system. The excess kurtosis is measured in the phase quadrature and its negative values indicate sub-Gaussian statistics, i.e.~lighter tails. Both quantities are non-Gaussian features in the state, induced by photon subtraction. All modes in the initial squeezed vacuum $V_0$ are equally squeezed (i.e.~$s_1= \dots =s_m$) at $10{\rm dB}$. \label{fig:BigGraph}}
\end{figure}

For larger graphs, it quickly becomes impractical to use visualisations of single-mode Wigner functions to grasp the properties of the state. Therefore, we must resort to more coarse grained quantities, such as the previously introduced excess kurtosis, to elucidate the non-Gaussian effects in larger graphs. In Fig.~\ref{fig:BigGraph} we consider photon subtraction from a large triangular graph, and evaluate the excess kurtosis for each vertex (bottom panel). A spread of non-Gaussian features throughout the graph is clearly visible, with the strongest effect at the vertex in which the photon was subtracted. As another perfect illustration of our main result (\ref{eq:Main}), we observe that the vertices which are more than two steps removed from the subtraction point remain completely unaffected.

Furthermore, it is well-established \cite{PhysRevA.86.012328,takahashi_entanglement_2010,ourjoumtsev_increasing_2007,walschaers_entanglement_2017,walschaers_statistical_2017} that photon addition and subtraction increase the entanglement between optical modes. Because the graph states that are studied in this letter are pure states, we can resort to rather straightforward entanglement measures. We limit ourselves to the study of bipartite entanglement, where we quantify the amount of entanglement between a single vertex and the remainder of the graph state. This entanglement can be measured using the purity of the single-mode state, associated with the vertex. This purity $\mu$ can be obtained from the Wigner function (\ref{eq:redWigner}) via
\begin{equation}\label{eq:purity}
\mu = 4\pi\int_{\mathbb{R}^{2}} {\rm d}^2\beta\, \abs{W_{\{k\}}(\beta)}^2.
\end{equation}
We then compare this purity to the purity $\mu_{\rm Gauss} = {\det V_{\{k\}}}^{-1/2}$ of the same vertex prior to photon addition or subtraction. Hence, when we obtain a relative purity $\mu/\mu_{\rm Gauss} < 1$, this serves as an indicator for an increase of entanglement due to photon subtraction or addition. In the top panel of Fig.~\ref{fig:BigGraph}, we clearly see that photon subtraction increases the entanglement in the graph. The enhancement is most profound in the vertex where the subtraction takes place. This is in agreement with the cruder study of entanglement in \cite{walschaers_entanglement_2017}, where we highlight that photon addition and subtraction enhance entanglement. Fig.~\ref{fig:BigGraph} sketches a more refined picture of the range of these effects for graph states. In particular, we see that entanglement properties of all vertices up to two steps removed from the point of subtraction are altered. Beyond these vertices, the entanglement properties remain unchanged, in accordance with (\ref{eq:Main}). Note that more subtle notions of multipartite entanglement cannot be studied as straightforwardly and require a more intricate framework \cite{PhysRevLett.92.087902,MINTERT2005207,PhysRevLett.111.110503}, which lies beyond the scope of this letter.\\


{\em In summary}, we showed that single-photon addition or subtraction in one vertex of a CV graph state only locally alters the state. Local is here understood in terms of neighbourhoods of the vertex where the addition or subtraction takes place. In particular, we demonstrated that only the vertex of subtraction itself, the vertices directly connected to it by an edge, and the vertices which are in turn connected to them are affected. In other words, non-Gaussian effect spread exactly two steps along the graph. This is confirmed in the examples of Figs.~\ref{fig:GraphWig} and \ref{fig:BigGraph}. In Fig.~\ref{fig:3}, we show that non-Gaussian effects can be spread out over a larger region by adding or subtracting the photon in a superposition of vertices. This further reaching non-Gaussianity comes at the price of being less pronounced in the affected vertices.

{\em As a future perspective,} in the context of quantum computational tasks, this result implies that non-Gaussian features can be introduced in a local way. 
Hence, the addition and subtraction of multiple photons in distinct vertices can be controlled  independently when these vertices are sufficiently far apart, i.e.~when the vertices that are affected by each addition or subtraction do not overlap.
On the other hand, if only weak non-Gaussian effects are required in every vertex, it is more convenient to use single-photon addition or subtraction in a superposition of vertices, as shown by Fig.~\ref{fig:3}. 

Nevertheless, it is important to recognise that the method (\ref{eq:clusterV}) to construct clusters through the application of $C_z$ gates is an idealised way of generating a CV graph state. In particular, the fact that $C_z$ gates also introduce squeezing implies a challenge for direct experimental implementations. State-of-the-art experiments typically start from a set of squeezed modes and apply a passive linear optics transformation to create an {\em approximate graph state} \cite{PhysRevA.76.032321,PhysRevA.91.032314,cai_reconfigurable_2016}, which are only equivalent to ideal graph states (\ref{eq:clusterV}) in the limit of infinite squeezing. These approximate graph states have weak correlations between vertices that are not connected by edges of the graph. Hence, non-Gaussian effects can spread further through approximate graph states than predicted by (\ref{eq:Main}). These effects must be taken into account when quantum computational protocols are implemented. How these non-Gaussian features, induced by photon addition or subtraction, can be harnessed to achieve a quantum computational advantage (let alone in practical quantum algorithms) now imposes itself as an open question.

Finally, our main result, the next-to-nearest neighbour spreading of non-Gaussian features, seemingly defies the local structure of the $C_Z$ gates, which suggests only nearest-neighbour effects. Due to the conditional nature of the photon addition or subtraction, arguments based on the no-signalling theorem are no longer valid in this context. Hence, our findings emphasise the need for a deeper mathematical understanding of conditional quantum operations.

\begin{acknowledgements} This work is supported by the French National Research Agency projects COMB and SPOCQ, and the European Union Grant QCUMbER (no. 665148). N.T. acknowledges financial support of the Institut Universitaire de France. M.W. is funded through Research Fellowship WA 3969/2-1 of the German Research Foundation (DFG).
\end{acknowledgements}

\bibliography{Paper_Clusters.bib}

\end{document}